\documentstyle[preprint,prd,aps]{revtex}
\begin{document}
\draft
\preprint{
\vbox{
\halign{&##\hfil\cr
	& AS-ITP-98-08 \cr
	& Sept. 1998 \cr}}
}
\title{Possible Effects of Quantum Mechanics Violation
Induced by Certain Quantum Gravity on Neutrino Oscillations}

\author{Chao-Hsi Chang$^{1,3}$, Wu-Sheng Dai$^4$, Xue-Qian Li$^{1,3,4}$,
Yong Liu $^5$, Feng-Cai Ma$^6$ and Zhi-jian Tao$^{1,2,3}$} 
\address{ $^1$CCAST (World Laboratory), P.O.Box 8730, Beijing 100080, China}
\address{ $^2$Institute of High Energy Physics, Academia Sinica, 
Beijing 100039, China.}
\address{ $^3$Institute of Theoretical Physics, 
 Academia Sinica, P.O.Box 2735, Beijing 100080, China}
\address{ $^4$Department of Physics, Nankai University, Tianjin, 300071, China}
\address{ $^5$Nankai Institute of Mathematics, Nankai University, 
Tianjin, 300071, China}
\address{ $^6$Department of Physics, Liaoning University, 
Shenyang 110036, China}

\maketitle
\vspace{-5mm}

\begin{abstract}

In this work we tried extensively to apply the EHNS postulation 
about the quantum mechanics violation effects induced by 
the quantum gravity of black holes to neutrino oscillations. 
The possibilities for observing such effects 
in the neutrino experiments (in progress and/or
accessible in the near future) were discussed. 
Of them, an interesting one was outlined specially.

\vspace{4mm}
{\bf PACS: 14.60.Pq, 03.75.-b, 03.65.Bz}
\end{abstract} 

\vfill \eject


\vspace{15mm}

It is known that neutrino oscillation is a 
very possible solution for the long-standing puzzle
about the `deficiency' of solar neutrino\cite{sun}
and the atmospheric
neutrino problem\cite{atom}. The 
neutrino oscillation experiments are very difficult
but people have been achieving progress steadly year to year. 
Especially, Super-Kamiokande has a high rate to collect 
the solar and atmospheric neutrino events. 
Recently they have reported some evidence for
neutrino oscillations\cite{superK}. 
In addition, several long-base neutrino
experiments for the oscillation in matter
are in progress as planned.

We also know that many years ago Hawking, based on 
the principle of quantum mechanics and gravity, proposed 
a very interesting conjecture that the quantum gravity effects 
of black holes may cause to emit particles in thermal spectrum\cite{Haw}. 
According to the conjecture, black holes may create particles in
pairs and some of particles fall back into the black holes while 
some of the others escape `away' thermodynamically, 
thus part of the information about the state 
of the system may be lost to the black holes. 
For a quantum mechanical system, due to the effects
caused by the microscopic real and virtual black holes,
the system which is in a pure quantum state may transmit 
to a mixed one, i.e. it manifests quantum mechanics violation (QMV). 
To describe a mixed quantum system from a pure state to a mixed one, 
instead of the wave function, density matrix description has to
be adopted\cite{Haw1}.
In such an evolution, where the QMV effects are involved,
CP, and probably CPT, can be violated
due to the non-local quantum gravity effects.
Thus Hawking's suggestion not only in a macroscopic 
configuration but also in a microscopic 
`elementary' particle level has received careful considerations. 
First of all, more than 10 years ago
Ellis, Hagelin, Nanopoulos and Srednicki (EHNS) \cite{Ellis},
being motivated to track the `sources' for CP and CPT violation, proposed
to observe the QMV induced by the quantum gravity effects in $K^0-\bar{K^0}$
system with additional reasonable assumptions, 
and then the authors of \cite{Huet,Ellis1} 
reexamed and formulated the effects with more care, and gave 
refreshed bounds on the parameters of the effects for QMV. To play the same
game, the authors of \cite{Du} discussed the possibility of observing
the effects in $B^0-\bar {B^0}$ system. The authors of \cite{Liu}
studied such effects on neutrino oscillation. 
As the kinetic energies and the masses stand on the same foot
for gravity in the stress-energy tensor, and no matter 
neutrino masses are zero or finite (a tiny nonzero mass at most) 
thus the energy of the neutrinos will play the roles in the stress-energy 
tensor when considering gravity.
Considering the fact that the neutrino oscillation observations
may take place in very differnt and
long distances, even so long as the distance from the Sun to the Earth,
it may have some advantage in looking for the QMV effects, at least, we 
should consider them quantitatively. In this paper, we will extend
to apply the EHNS formulation to the neutrino cases,
and discuss the effects of the QMV affecting various neutrino oscillation
observations, relevant to the present and planned neutrino 
experiments.

To follow the notation of EHNS,
let us repeat briefly their formulation
so as to start the calculations and discussions here.

To describe such states, commonly instead of wave function, the
density matrix
is employed. The density matrix of a pure state can always be written as 
\begin{equation}
\rho _{pure}=|\psi ><\psi |,
\end{equation}
while a mixed state then should be in the form 
\begin{equation}
\rho _{mix}=\sum_aP_a|\psi _a><\psi _a|,\hspace{1cm}{\rm with}%
\;\;\sum_aP_a=1,
\end{equation}
where $|\psi >$ and $\psi _a>$ are the regular wave functions respecting the
superposition rule and normalization $<\psi |\psi >=1$, $<\psi _a|\psi _a>=1$
(not to sum over $a$). Note that 
\begin{equation}
Tr(\rho _{pure})=Tr(\rho _{mix})=1,
\end{equation}
but 
\begin{equation}
Tr(\rho _{pure}^2)=Tr(\rho _{pure})=1,\;\;\;Tr(\rho _{mix}^2)<1.
\end{equation}
The Schr\"odinger equation for the density matrix is accordingly written as 
\begin{equation}
\label{Schr}i{\frac \partial {\partial t}}\rho =\left[ H,\rho \right] ,
\end{equation}
where $\rho $ can be either $\rho _{pure}$ or $\rho _{mix}$. Indeed so far
it is exactly equivalent to the regular form of the Schr\"odinger equation
for the wave functions. It is easy to prove that with Eq.(\ref{Schr}) one has 
$$
{\frac d{dt}}Tr(\rho ^2)=0, 
$$
namely, pure and mixed states never interchange. However,
as EHNS suggested \cite{Ellis}, the Hawking's 
quantum gravity effects at vicinity of real and virtual black holes may 
violate quantum mechanics i.e. modify the Schr\"odinger equation 
significantly. For simplicity, we derive all formulae for a two-energy-level
system as an illustration.
Generalizing the Schr\"odinger equation of a two-energy-level system, one
can expand the $2\times 2$ matrix form of $\rho $ and $H$ in terms of $%
\sigma _0$ and $\sigma _i$, where $\sigma _0$ is a $2\times 2$ unit matrix
and $\sigma _i (i=1,2,3) $ are the well-known Pauli matrices, i.e. 
\begin{equation}
\label{Pauli} \rho =\rho _0\sigma _0+\rho _i\sigma _i,\;\;\;H=H_0\sigma
_0+H_i\sigma _i.
\end{equation}
Thus besides the trivial $\rho_0$ component, Eq.(\ref{Schr}) can be recast
into a tensor form as \cite{Huet} 
\begin{equation}
\label{exsch} i{\frac d{dt}}\rho =2\epsilon ^{ijk}H^i
\rho ^j\sigma ^k,\;\;\;(i,j,k=1,2,3).
\end{equation}
Due to the QMV effects being included by 
the concerned quantum gravity, EHNS introduced 
a non-hermitian piece to Eq.(\ref{exsch}), 
which modifies the Schr\"odinger equation Eq.(\ref{Schr}) greatly.
The newly additional non-hermitian term is 
\begin{equation}
\label{exsch1} i\delta H\rho =-h^{0j}\rho ^j\sigma _0-h^{j0}\rho ^0\sigma _j-h^{ij}\sigma
_i\rho ^j.
\end{equation}
Since probability is conserved, and 
its entropy should not decrease, it is required 
$$
h^{0j}=h^{j0}=0. 
$$

EHNS \cite{Ellis} and the authors of ref.\cite{Huet} applied
this modified Schr\"{o}dinger equation to the $K^{0}-\bar K^{0}$ system.
By enforcing different conservation
laws on the effects, $h^{ij}$ would be constrained. 
If a physical quantity is conserved, its corresponding operator 
$O$ must commute with the Hamiltonian and requires $d/dt(TrO\rho)=0$. 
Hence 
$$
Tr(O\delta H\rho)=0.%
$$

EHNS and the authors of \cite{Huet}
assumed $O=\sigma_1$ which corresponds to strangeness
being conserved ($\Delta S=0$) 
in the neutral kaon system:
$$
<K^0|\sigma_1|K^0>=-1,\;\;\;\; {\rm while}\;\;\;\; <\bar K^0|\sigma_1|\bar
K^0>=+1. 
$$
Then $h_{\mu\nu}$ of Eq.(8) can be written as
a $4\times 4$ matrix:
\begin{equation}
\label{str} h_{\mu\nu}=2\left( 
\begin{array}{cccc}
0 & 0 & 0 & 0 \\ 
0 & 0 & 0 & 0 \\ 
0 & 0 & -\alpha & -\beta \\ 
0 & 0 & -\beta & -\gamma 
\end{array}
\right). 
\end{equation}

Whereas EHNS also proposed an alternative parameter set by
assuming the conservation operator is $O=\sigma_3$, 
and it is the case that energy and the other
quantum mumbers such as leptonic number etc are conserved. 
Thus the matrix 
$h_{\mu\nu}$ reads 
\begin{equation}
\label{bary} h_{\mu\nu}=2\left( 
\begin{array}{cccc}
0 & 0 & 0 & 0 \\ 
0 & -\alpha & -\beta & 0 \\ 
0 & -\beta & -\gamma & 0 \\ 
0 & 0 & 0 & 0 
\end{array}
\right). 
\end{equation}

Note that here we have added an extra factor 2 in front of the matrix at
Eq.(\ref{bary}) which is a different parametrization from the notation given
in \cite{Ellis}, the reason is to make the form similar to that in Eq.(\ref
{str}) where the authors of \cite{Ellis} had put a factor 2 (see
Eqs.(2.31) and (3.15) in \cite{Ellis}). 

For the parametrization of Eq.(\ref{bary}), to avoid
the states with complex entropy,  $Tr\rho^2$ can never exceed
unity, so it requires 
$$
\rho_{\alpha}H_{\alpha\beta}\rho_{\beta}\leq 0,%
$$
thus
\begin{equation}
\label{condi} \alpha>0,\;\;\;\;\gamma>0,\;\;\;\;\alpha\gamma>\beta^2.%
\end{equation}

By fitting data of $\epsilon$ and the semileptonic decays of the
K-system, EHNS obtained \cite{Ellis}
$$
\alpha+\gamma\leq 2\times 10^{-21} \; {\rm GeV } ,%
$$
while Huet and Peskin \cite{Huet} updated the values as 
$$
\beta=(3.2\pm 2.9)\times 10^{-19}\;{\rm GeV},\;\;\;\; \gamma=(-0.2\pm
2.2)\times 10^{-21}\; {\rm GeV}. 
$$
and recently Ellis et al. gave a further estimate as \cite{Ellis1} 
\begin{equation}
\label{new} \alpha\leq 4\times 10^{-17} \;{\rm GeV},\;\;\;\; |\beta|\leq
3\times 10^{-19}\;{\rm GeV},\;\;\;\; \gamma\leq 7\times 10^{-21}\;{\rm GeV}. 
\end{equation}
Indeed, the only important issue quoted here is the order of 
magnitudes of $
(\alpha,\beta,\gamma)$ and the concrete coefficients are not much of
significance.

Since the QMV effects are caused by quantum gravity, it is suggested
that $\alpha,\;\beta,\;\gamma$ be proportional to $M^2/M_{pl}$ where $M_{pl}$
is the Plank mass and $M$ is an energy or mass scale of the concerned
physical process occurring in our quantum system (neutral kaon or neutrino
under consideration).

Now let us turn to the case for neutrino
oscillations.

The quantum gravity effects must play similar roles in all the
quantum systems as that in the $K^0-\bar K^0$ system, but the crucial
problem is if they are observable or not. The neutrino oscillations 
among different species neutrinos will be affected by the 
concerned QMV effects, and might be observable. Because 
we may observe the oscillations at very different distances 
in `vacuum' and in matter as well, one may expect 
to have more advantages for observing such effects in 
neutrino oscillation systems
than in the $K^0-\bar{K^0}$ system or else.
In the two-generation neutrino oscillation
case, the $h_{\mu \nu }$ has just the form as Eq.(\ref{Pauli})
in the basis of  Pauli matrices, whereas, in three-generation
neutrino oscillation case, it becomes more complicated
while the Pauli matrices
will be replaced by the $SU(3)$ Gell-Mann matrices\cite{clt}.
In this work, as stated above, for simplicity we will restrain
ourselves only to formulate the two-generation
neutrino case. Namely we will only consider 
the form of $\delta H$ given in Eq.(\ref{str}) and Eq.(\ref{bary}). 
It certainly is interesting to note here that besides those
expected effects, the lepton number is allowed 
to violate\footnote{Here the lepton
number is violated due to the interaction of the black holes.} even if 
the neutrinos are massless. Furthermore, one will
see that the effects themselves may induce
oscillations so they are observable
in present or accessible 
neutrino oscillation experiments.

The physical picture may be imagined as the following.
As neutrinos $\nu_i$ interact with the heavy object, a `micro black hole', 
due to the quantum effects,
the black hole creates a pair of neutrino-antineutrino of certain species,
the neutrino-antineutrino pair interacts with
the coming neutrino in a certain (coherent or incoherent) manner, 
afterwards, a neutrino and a antineutrino fall into the black hole
but one neutrino may escape away to respect the coming neutrino.
Whereas we should note that the escaping neutrino does not need to be the
same as the coming one. Which one escapes, only  
depending on its coupling to the micro black hole via gravity,
namely the two species (of course may be the same) neutrinos have different
couplings to the black holes (if $a\neq b$).
If $\alpha\neq\gamma$ in the QMV terms,
$h_{ij}$ of Eq.(9) and Eq.(10) indeed reflects such facts.

Thus we may try to apply this scenario to neutrino oscillation \cite{Liu}. 
In order to have some idea about magnitude order of the effects
for definiteness and comparison with the $K^0-\bar {K^0}$ system
as possible as one can,
we further try to assume the corresponding parameters for neutrino
systems as the follows:
\begin{equation}
\label{theo} \alpha_{\nu}\leq 4\times 10^{-17} ({\frac{E_{\nu}}{0.5}})^2  \;%
,\;\;\;\; |\beta_{\nu}|\leq 3\times 10^{-19}({\frac{E_{\nu}}{0.5}}%
)^2\;,\;\;\;\; \gamma_{\nu}\leq 7\times 10^{-21}({\frac{E_{\nu}}{0.5%
}})^2\;, 
\end{equation}
where $E_{\nu}$ is the energy of the emitted neutrino, and every
quantities in the above are in GeV. Here 0.5 corresponds to
the mass of kaon. 

Note again that the parameters can be very different from that listed above,
but we just assume them as a reference for later discussions.
If assuming the solar neutrino deficit is due to neutrino
oscillation \cite{Mik}, the parameter set Eq.(\ref{theo}) will be 
restricted by data. Later we will show that the solar
neutrino and other neutrino experiments on the Earth may set some 
substantial constraints on the parameters.

Now let us discuss the meaning of the solutions obtained from 
Eq.(\ref{str}) and Eq.(\ref{bary}).

\noindent
{\bf I. The asymptotic behavior of QMV evolution of the
neutrino system}

It is easy to realize that the expressions Eq.(\ref{str}) and 
Eq.(\ref{bary}) would lead
to different behaviors for the neutrino oscillations. 

{\sf a).} With the form of $\delta H$ given in Eq.(\ref{str}), one 
has the solution that a probability
for $\nu _e$ transition to another $\nu _x$ in vacuum ($x$ can be $\mu ,\tau 
$ or a sterile neutrino flavor), 
\begin{equation}
\label{str1}P_{\nu _e\rightarrow \nu _x}={\frac 12}-{\frac 12}e^{-\gamma
L}\cos {}^22\theta _v-{\frac 12}e^{-\alpha L}\sin {}^22\theta _v\cos ({\frac
\Delta {2E_\nu }}L),
\end{equation}
where $\Delta \equiv |m_{\nu _1}^2-m_{\nu _2}^2|$, L is the distance from
the production spot of $\nu _e$ to the detector and $\theta _v$ is the mixing
angle of $\nu _e$ and $\nu _\mu $ in vacuum. To obtain the above formula,
one should assume $\beta \ll \alpha ,\gamma $, in fact this approximation is
not necessary, but here only for demonstration convenience, otherwise the 
formula would become tedious. In the work\cite{Liu}, more precise
numerical results were given.

It is noted that in the basis of mass, because 
$|\nu_e>=(\cos\theta_v|\nu_1>
+\sin\theta_v|\nu_2>)$ and $|\nu_{\mu}>=
(-\sin\theta_v|\nu_1>+\cos\theta_v|\nu_2>)$, so
$$<\nu_e|\sigma_1|\nu_e>=2\sin\theta_v\cos\theta_v,\;\;\;\;
<\nu_{\mu}|\sigma_1|\nu_{\mu}>=-2\sin\theta_v\cos\theta_v. $$
As the case of $K^0-\bar {K^0}$ system, the conservation of 
$\sigma_1$ should mean that flavor conserves
and there would be no transition among different flavors.
At the first glimpse, $\delta H$ seems cause a flavor transition.
In fact, if the original Hamiltonian does mix the flavors for
massive neutrinos, the $\delta H\rho$ term does not cause it
further, but strengthens or weakens the transition caused by 
the original Hamiltonian only. One can see that in the 
case an exponential factor exists in front of the 
harmonic oscillation, which is our familiar expression of 
neutrino oscillation in vacuum. Thus this
extra factor changes the oscillation behavior, but does not cause it.

When the neutrinos are massless, it is another story. Then
the mixing disappears, i.e. $\theta_v=0$,
then $<\nu_e|\sigma_1|\nu_e>=<\nu_{\mu}|\sigma_1|\nu_{\mu}>=0$. 
It implies that the two states are degenerate in the 
regular QM framework. But as long as there are extra terms 
such as the QMV, the degeneracy is broken
and an oscillation can occur due to the new effects. Hence in 
this case, the $\sigma_1$ conservation does not forbid 
such a transition 
between the different flavors, (because expectation value of
$\sigma_1$ is zero for all flavors).

In fact, if neutrinos are massless, the oscillation is still expected. 
Namely if considering Eq.(\ref{exsch}) only, 
we have $m_{\nu}=0$, and $\nu_1,\nu_2$ are exactly $\nu_e,
\nu_{\mu}$, but with the extra $\delta H\rho$ Eq.(\ref{exsch1})
in the evolution equation, and different couplings being indicated as
$h_{\mu\nu}$ in Eq.(9), the difference
for different `flavor' neutrinos
manifests. Thus $\nu_1$ and $\nu_2$ have different 
behaviors as they propagate in an environment full
with the micro black holes and `oscillations' between them appear.

{\sf b).} With expression Eq.(\ref{bary}), we have the solution\cite{Chang}: 
\begin{equation}
\label{bary1} P_{\nu_e\rightarrow\nu_{x}}={\frac{1}{2}}\sin^2
2\theta_v\left( 1- e^{-(\alpha+\gamma)L}\cos({\frac{\Delta}{2E_{\nu}}}L)
\right). 
\end{equation}
Note that to obtain the above result, we assumed that $(\alpha,\beta,\gamma)_{\nu} \ll{
\frac{\Delta}{2E_{\nu}}}$. 

In fact, the exact result depends on the fact if the factor $\kappa^2$
is greater, equal or smaller than zero with the definition
\begin{equation}
\label{det} \kappa^2\equiv 4\left[(\alpha-\gamma)^2+4\beta^2-{\frac{\Delta^2
}{4E^2_{\nu}}} \right]. 
\end{equation}
If $\kappa^2$ is less than zero, the
oscillating form of Eq.(\ref{bary1}) is resulted in, 
only when $\kappa^2$ is greater or equal to zero, 
the expression turns into a purely damping
solution. The precise version of Eq.(\ref{bary1}) is
\begin{equation}
\label{full} P_{\nu_e\rightarrow\nu_{x}}={\frac{1}{2}}\sin^2
2\theta_v \{1-e^{-(\alpha+\gamma) L}\left[{\frac{\alpha-\gamma}{\kappa}}%
(e^{\kappa L/2}-e^{-\kappa L/2}) +{\frac{1}{2}}(e^{\kappa L/2}+e^{-\kappa
L/2})\right] \}. 
\end{equation}
Indeed when $\kappa^2<0$, $\kappa$ is imaginary, the solution 
contains an oscillatory factor, otherwise attenuative.

Let us discuss the phenomenological significance of Eq.(\ref{str1}) and 
Eq.(\ref{bary1}) in the below.

\noindent
{\bf II. The equation Eq.(\ref{str1}) and Eq.(\ref{bary1})
leads to completely different asymptotic limits as $r\rightarrow \infty$
(or $L\rightarrow\infty$)}

The exponentially damping term in Eq.(\ref{str1}) would wash out any
information of neutrino mixing as long as the detector is placed far enough
from the source. In that case, $P_{\nu_e\rightarrow\nu_{\mu}}(t\rightarrow%
\infty)= {\frac{1}{2}}$ for two generations, and if generalizing
the result to the n-generation structure\cite{Liu}:  
$$
P_{\nu_e\leftrightarrow\nu_{\mu}}(t\rightarrow\infty)={\frac{1}{n}},\;\;
P_{\nu_e\leftrightarrow\nu_{\tau}}(t\rightarrow\infty)={\frac{1}{n}},\;\;
\;\; P_{\nu_{\tau}\leftrightarrow\nu_{\mu}}(t\rightarrow\infty)={\frac{1}{n}}%
.%
$$
On the contrary, Eq.(\ref{bary1}) would lead to a different consequence as 
$$
P_{\nu_e\rightarrow\nu_{\mu}}(t\rightarrow\infty)={\frac{1}{2}}
\sin^22\theta_v ,%
$$
i.e. the mixing angle between $\nu_e$ and $\nu_{\mu}$ is still
there; for the 3-generation case we will
have a similar result, only the simple `Cabibbo-like' angle $%
\theta_v$ should be replaced by the `KM-like' entries. 

All the above expressions can apply to the $\nu_a\leftrightarrow \nu_b$ case
with a,b being any pair of $e,\mu,\tau$ as long as $a\neq b$.

\noindent
{\bf III. The solar neutrino problem vs. the QMV effects}

{\sf a).} For $\Delta\simeq 10^{-5}{\rm eV}^2$, i.e. MSW solution 
for the solar neutrino puzzle, one expects the averaged effect of 
oscillation term $\cos(\frac{\Delta}{2E_{\nu}}L)$ vanishes. 
Therefore the transitional probability
can be re-written as the follows:

In the case of Eq.(\ref{str}) 
\begin{equation}
\label{MSW} P(\nu_e\to\nu_e)=\frac{1}{2}[1+(1-2X)e^{-\gamma L}\cos 2\tilde{%
\theta}_0 \cos 2\theta]. 
\end{equation}
where $\tilde{\theta}_0$ is the neutrino mixing angle in the center of the
Sun. $X$ is the jumping probability from one neutrino mass eigenstate to
another in the MSW resonant region. For the large angle solution $X\simeq 0$
and for the non-adiabatic solution it can be close to one. From 
Eq.({\ref{MSW}})
we may see that $e^{-\gamma L}>>1$ is not favored to fit the solar 
neutrino data besides violating the condition Eq.(\ref{condi}). 
Because in this case we obtain a constant suppression 0.5, which is
disfavored\cite{Kra}. As the result the bound $\gamma L\le 1$ is enforced. If
$\gamma L<<1$, the new violation effect is negligible. So only for $\gamma
L\sim O(1)$ the MSW solution for the solar neutrino problem should be modified.
Here $L$ is the distance between the Sun and the Earth. Generally we get $%
\gamma\leq 6\times 10^{-9}{\rm km}^{-1}$.

In the case of Eq.(\ref{bary}), the new effects are 
averaged to be zero over the
distance $L$. The situation is exactly the same as the MSW solution  without
the QMV terms. In this case one cannot obtain any information
about the QMV from fitting the solar neutrino data.

{\sf b).} For the vacuum oscillation solution to the solar neutrino problem, $%
\Delta \sim 10^{-10}{\rm eV}^2$. In this case the oscillation term is not
averaged to be zero. The transitional probability is given in Eq.(\ref{str1}), 
Eq.(\ref{bary1}) and Eq.(\ref{full}). Again we obtain the 
bound $\alpha L, \gamma
L\leq 1$ in order to solve the solar neutrino puzzle. And only for $\alpha
L, \gamma L\sim 1$ the parameter region of the vacuum oscillation solution
should be modified.

\noindent
{\bf IV. A very interesting feature indicated by Eq.(\ref{str1}) } 

Even if neutrinos are massless, the micro black hole effects can
induce neutrino transition from one flavor state to another.
For an $n$-flavor neutrino case, 
the oscillation probability could be simplified as
\begin{equation}\begin{array}{l}
\label{str2} 
P_{\nu_e\rightarrow\nu_x}={\frac{1}{n}}-{\frac{1}{n}}%
e^{-\gamma L} \\
P_{\nu_e\rightarrow\nu_e}={\frac{1}{n}}+{\frac{n-1}{n}}%
e^{-\gamma L} \\
\end{array}
\end{equation}
where SU(n) should replace SU(2) for the case of two neutrino species.
 
Indeed it is interesting to `check' if this oscillation 
probability alone is enough to solve the solar neutrino
problem without requiring nonzero neutrino mass.

First of all, it is realized that $\gamma$ cannot be a constant, otherwise
the $\nu_e$ suppression is energy independent which disagrees with the
solar neutrino data \cite{Kra}. By dimension analysis, one may assume $%
\gamma=\gamma_0E^2_{\nu}\sim \frac{E_{\nu}^2}{M_{pl}}$ for massless
neutrino. With this assumption we see that the larger neutrino energy
corresponds to larger suppression. So in the solar neutrino experiment the $%
^8$B neutrino is suppressed most, which is $\frac{1}{n}$. $^7$Be neutrino
are suppressed less but very close to $\frac{1}{n}$. pp neutrino is
suppressed least which is between $\frac{1}{n}$ and 1. After careful study
we find that the solar neutrino data can be fitted best with $n=3$. So in
the following we will discuss three species neutrino case. We adopt the
standard solar model (BP98) \cite{BP98} for our discussions.
The predicted neutrino flux for $ H_2O$ experiment is:  
for $\nu_e\to\nu_{\mu},\nu_{\tau}$ oscillation 
\begin{equation}
\Phi_{H_2O}^{th}=\left( 2.21
\begin{array}{c}+0.19 \\ -0.14\end{array} \right) \times 10^6{\rm cm}^{-2} {\rm 
s%
}^{-1} 
\end{equation}
for $\nu_e\to\nu_{\mu},\nu_s$ oscillation ($\nu_s$ is a sterile neutrino) 
\begin{equation}
\Phi_{H_2O}^{th}=\left( 2.01\begin{array}{c}+0.19\\-0.14\end{array}\right)
\times 10^6{\rm cm}^{-2}
s^{-1} {}
\end{equation}

The observed flux $\Phi_{H_2O}^{exp}$ is 
$2.42\pm 0.06\begin{array}{c} +0.10\\ -0.07 \end{array}%
\times 10^{6} {\rm cm}^{-2}{\rm s}^{-1}$ from 
Super-Kamiokande\cite{superK} (it is $2.80\pm 0.19\pm 
0.33\times 10^{6} {\rm cm}^{-2}{\rm s}^{-1}$ from Kamiokande). 
The ratio $\Phi_{H_2O}^{th}/\Phi_{H_2O}^{exp}$
is estimated to be $0.92\pm 0.17$ and $0.84\pm 0.17$ for $%
\nu_e\to\nu_{\mu},\nu_{\tau}$, and $\nu_e\to\nu_{\mu},\nu_s$ oscillation
respectively. Theory agrees with the experiment within $1\sigma$.

The neutrino capture rate in the chlorine experiments is obtained as 
\begin{equation}
S_{Cl}^{th}=2.6\pm 0.4
~~~~{\rm SNU}
\end{equation}
compared with the observed one $S_{Cl}^{th}=2.55\pm 0.25~~{\rm SNU}$
\cite{Bah1}. The ratio is estimated as 
\begin{equation}
S_{Cl}^{th}/S_{Cl}^{exp}=1.0\pm 0.1 
\end{equation}
So the theoretical expectation is in very good agreement with the experiment.

For the gallium experiments if the parameter $\gamma_0$ falls in the region 
$(1.5-3.7)\times 10^{-8}{\rm MeV}^{-2}{\rm km}^{-1}$, we obtain the capture
rate as 
\begin{equation}
S_{Ga}^{th}=(68 \sim 79)\;\; {\rm SNU} 
\end{equation}
This agrees with the experimental value $(73.4\pm 5.7) {\rm SNU}$\cite{Bah1} at 
$1\sigma$ level. At $2\sigma$ level $\gamma_0$ can be taken a value from $%
(0.66-5.4)\times 10^{-8}{\rm MeV}^{-2}{\rm km}^{-1}$.

\noindent
{\bf V. The scenario for the `atmospheric neutrino' problem} 

With three-generation neutrinos and $\gamma _0$ given in {\bf IV},
by fitting solar neutrino data at the level
of $2\sigma $ errors, we may estimate some of the observables
further for the atmospheric neutrino observations. 

The up-down asymmetry of $\mu$-like and e-like
events for $\cos \Theta >0.2$ (down) and $\cos \Theta <-0.2$ (up), where 
$\Theta $ is zenith angle, are denoted by $Y_e,Y_\mu $. In the present 
scenario $Y_e$ is always close to $1$, independent of the energy
and the traveling distance of the neutrinos, it is in agreement 
with the data $Y_e({\rm sub-GeV})=1.13\pm 0.08$ and 
$Y_e({\rm multi-GeV})=0.83\pm 0.13$ \cite{superK}. However 
$Y_\mu $ is estimated as 0.62-0.98, 0.5 for multi-GeV and sub-GeV events
respectively, i.e. at $1\sigma $ level agrees with the measured 
$Y_\mu$(multi-GeV)$=0.54\pm 0.07$, but at $4\sigma $ level
with the measured $Y_\mu $(sub-GeV)$=0.78\pm 0.06$. It is a 
little more involved to calculate the ratio of the
total $\mu$-like and e-like events. Here we only give a rough estimate
on the double ratio by approximating that the down-going neutrino
flux is almost unsuppressed, while the up-going and the horizontal neutrino
flux is suppressed by a factor of $1/3$. They are estimated 
to be 0.6 and 0.5-0.6 for sub-GeV and multi-GeV events respectively
with the scenario. This is also not bad but in agreement with
the measured ones $0.61\pm 0.03\pm 0.05$, $0.66\pm 0.06\pm 0.08$\cite{superK}.
We conclude here that our scenario with zero neutrino mass can fit all the
measurements of the solar and atmospheric neutrino experiments, except 
$Y_\mu $(sub-GeV).

{\bf VI. An alternative mechanism for the solar
and atmospheric neutrino flux `shortage'} 

As estimated in the previous sections, the QMV effects
can serve as an alternative mechanism for the solar
and atmospheric neutrino flux shortage, if assuming 
\begin{equation}
\label{facto} (\alpha,\beta,\gamma)_{\nu}=(\alpha,\beta,\gamma)_{K}\cdot
({\frac{E_{\nu}}{M_K}})^2.  
\end{equation}
The factor $(E_{\nu}/M_K)^2$ appears here is due to
that we think that there is some relation between neutrino oscillation
and $K^0-\bar{K^0}$ system for QMV induced by quantum gravity.
Namely based on the ansatz proposed by EHNS, we will have 
$(\alpha,\beta,\gamma)_K\propto M^2/M_{pl}$
for the $K^0-\bar{K^0}$ system, and a similar 
parametrization for neutrinos 
$(\alpha,\beta,\gamma)_\nu\propto E_\nu^2/M_{pl}$
thus the factor 
$(E_{\nu}/M_K)^2$ appears in Eq(\ref{facto}). 
In the solar neutrino case, 
$E_{\nu}\sim 0.3-10$ MeV, the factor suppresses 
$(\alpha,\beta,\gamma)_{\nu}$
by a factor of order $10^{-6}-10^{-4}$. This postulation should be tested by
experiments on the Earth.

The data on $\nu _\mu \rightarrow \nu _\tau $ oscillation by the CHARM II 
\cite{CHARM} collaboration claimed that no evidence of the neutrino flux
change had been observed. In the experiment, $E_\nu \sim 27$ GeV and $L\sim
0.6$ km. Considering the errors, $|\alpha L|$ must be smaller than 
$10^{-3}$. This constraints requires 
\begin{equation}
\label{limit}(\alpha _0,\beta _0,\gamma _0)_\nu \leq 2\times 10^{-12}{\rm MeV%
}^{-2}{\rm km}^{-1}.
\end{equation}
Combining Eq.(\ref{limit}) with the enhancement factor 
$({\frac{E_\nu }{M_K}})^2\sim 1.6\times 10^3$, it indicates 
$$
(\alpha ,\beta ,\gamma )_K\leq 4\times 10^{-24}{\rm GeV}. 
$$
This number is much below the upper bounds given by the authors of \cite
{Huet,Ellis1}. If this is the case, the violation effects of the quantum
mechanics would hardly influence the $\epsilon -$value in the neutral kaon
system. This constraint also excludes the $\nu _e,\nu _\mu ,\mu _\tau $
oscillations discussed in {\bf IV.} and {\bf V.}. Hence a sterile 
neutrino must be introduced and $\tau$ neutrino must be treated 
differently from the other species.

However, as pointed out above, the parameters for neutrino system do not
need to be the same as that for neutral kaon system, so this comparison
has qualitative meaning only.

\noindent
{\bf VII. The parameters $(\alpha,\beta,\gamma)_{\nu}$}

If we assume the parameters obtained in the neutral kaon system can be
generalized to the neutrino system through certain relation, then 
besides Eq.(\ref{facto}), we can have
the following possibilities.

A possible adoption could be that 
\begin{equation}
\label{mm} (\alpha,\beta,\gamma)_{\nu}\propto ({\frac{m_{\nu}}{M_K}})^2,
\end{equation}
whereas if $m_{\nu}$ is in the order of magnitude about a few tens 
of eV (for the $\tau$ neutrino probably), it is easy to show that 
in neutrino systems such an `adoption' would kill any possible
observational effects of the QMV induced by the micro black holes.

The scenario described by Hawking is that the quantum effects cause 
the micro black holes to radiate particle pairs with one of the pair 
falling into the event horizon while the other escaping
away. Considering this picture, an alternative postulation 
\begin{equation}
\label{Em} (\alpha,\beta,\gamma)\propto {\frac{E_{\nu}\cdot m_{\nu}}{M_{pl}}}, 
\end{equation}
could be reasonable if there is a nonzero neutrino mass, because the
escaping particle is moving relatively to the black holes. 

Supposing $m_{\nu_{\tau}}\sim 10$ eV, with the condition of the CERN-SPS
wide band neutrino beam (WBB) \cite{CHARM} the ansatz Eq.(\ref{Em}) 
results in 
$$
(\alpha,\beta,\gamma)_{\nu}\sim {\frac{27\times 10\times 10^{-9}}{(0.5)^2}}
\times (\alpha,\beta,\gamma)_K\simeq 6\times 10^{-27}- 4\times 10^{-23}\;%
{\rm GeV},%
$$
in terms of the values given in \cite{Ellis1,Liu}. If it is the case, the
values of $(\alpha,\beta,\gamma)_{\nu}$ satisfy the condition Eq.(\ref{limit})
set by the CHARM II data. Similar to what we did 
in in {\bf IV.} and {\bf V.} we can 
also fit the solar and atmospheric neutrino data in this case, most of 
the results are approximately the same except $\gamma_0
\simeq 10^{-9}\sim 10^{-10}{\rm MeV}^{-1}
{\rm km}^{-1}$. Here $\gamma\equiv\gamma_0E$.

Therefore such phenomena may be accessible 
in the proposed long-baseline neutrino oscillation experiments
where the neutrinos propagate sufficiently far to make the
damping effects observable. 

Let us give numerical estimation for the proposed 
long-baseline neutrino oscillation experiments.

\noindent
{\bf VIII. The scenario in long-baseline experiments}

In the planned long-baseline experiments,
KEK-Super-Kamiokande (250 km), CERN-GranSasso (730 km) and Fermilab-Sudan II
(730 km), the average energies of the $\nu_\mu$ beams are
approximately $1$ GeV, $6$ GeV and $10$ GeV \cite{Hir}. Accordingly, the
suppression factor $e^{-\gamma L}$ for the experiments should be 
\begin{equation}
0.2,\;\;\;\sim 0.0,\;\;\;\sim 0.0,
\end{equation}
respectively. These factors are estimated based on the value of $\gamma $
obtained above for the $E_\nu ^2$ dependence postulation Eq.(\ref{facto}). 
So the QMV effects should be observable in these long-baseline
experiments or the observation of the effects in the experiments
will make more stringent constraints on the parameters.

For the $E_\nu$ dependence postulation  Eq.(\ref{Em}) the
damping effect 
is negligibly small, so that the planned long-baseline experiments are 
unable to observe the QMV effects.

\noindent
{\bf IX. Summary}

It is an interesting subject for both aspects: first, the
conclusion would indicate, even indirectly, if there are
the mysterious micro black hole effects, secondly, if this picture 
is valid, their existence may be non-negligible in certain physical 
processes, especially, the neutrino oscillations would be affected and 
the resultant neutrino-flux attenuation may become observable at the 
planned long-baseline experiments.

Moreover, as we indicated above, if $(\alpha,\beta,\gamma)_{\nu}> 
{\frac{\Delta}{4E_{\nu}}}$ in Eq.(\ref{det}), the harmonic oscillation 
form would turn into a pure exponential damping form.

The ansatz Eq.(\ref{str}) and Eq.(\ref{bary}) lead to 
different asymptotic limits
as $t\rightarrow\infty$, for $P(\nu_a\rightarrow\nu_b)$ 
with $a\neq b$ being
certain species of neutrinos. If the distance between detector and source is
large enough, this difference of Eq.(\ref{str}) and Eq.(\ref{bary}) is
distinguishable.

If a sophisticated neutrino detector is located in Beijing to
receive neutrino flux sent from KEK, CERN and Fermilab, as suggested
by He and his collaborators \cite{He}, the
distance is remarkably large, then according to the above analysis,
observational prospect of such phenomena is optimistic.

In summary, neutrino oscillation
experiments may put a stronger bound on the QMV effects than the
K meson system if the $E_\nu ^2$ Eq.(\ref{facto}) 
dependence postulation is valid. Whereas in the case
of the $E_\nu$ dependence Eq.(\ref{Em}), K meson system would give
stronger restrictions. Even if neutrinos
are massless the QMV induced by micro black hole may `cause' neutrino
oscillation. We find that this oscillation has interesting prediction
for solar neutrino and atmospheric behavior. Moreover long-baseline
experiments on neutrino oscillation may provide us valuable information
about the neutrinos and the QMV effects.
Anyhow, the physical picture about the micro black holes has
phenomenological significance, especially to the neutrino 
oscillation problem. It is worth further and
deeper studies.

\vspace{4mm} 
{\bf Acknowledgment:} This work was supported in part by the
National Natural Science Foundation of China (NNSFC) and Grant No. LWTZ-1298
of the Chinese Academy of Sciences. One the authors (C.-H. Chang) would like
to thank JSPS for supporting him to visit Japan, and thank Prof. M. Kobayashi 
for warm hospitality during his visit of Institute of Particle and Nuclear 
Studies, KEK, Japan, as quite a lot of his part of the work is completed at 
KEK.

\end{document}